\documentclass{elsart}

\usepackage{graphicx}
\usepackage{amssymb}
\usepackage{amsfonts}
\usepackage{mathrsfs}
\usepackage{MnSymbol}
\usepackage{cite} %for references

\newcommand{\bs}[1]{\boldsymbol{#1}} % boldsymbol
\newcommand{\tr}[1]{\textrm{#1}} % roman text

\begin{document}
\begin{frontmatter}

%_______________________________________________________________
\title{Algebraic Gorkov solution in finite systems for the separable pairing interaction}

\author{R. M. Id Betan,$^{1,2,3}$ C. E. Repetto$^{1,2}$}
\address{$^{1}$Physics Institute of Rosario (CONICET-UNR), 
             	Esmeralda y Ocampo, S2000EZP Rosario, Argentina.}
\address{$^{2}$Department of Physics FCEIA (UNR),
             	Av.~Pellegrini 250, S2000BTP Rosario, Argentina.}
\address{$^{3}$Institute of Nuclear Studies and Ionizing Radiations (UNR), 
		    	Riobamba y Berutti, S2000EKA Rosario, Argentina.}

%__________
\date{\today}

%__________
\begin{abstract}
An algebraic Quantum Field Theory formulation of separable pairing interaction for spherical finite systems is presented. The Lipkin formulation of the model Hamiltonian and model wave function is used. The Green function technique is applied to obtain the model energy through the spectral function. Closed equation for the many-body energy of the system is given and comparison with exact models are performed. 
\end{abstract}

%___________

%____________
\end{frontmatter}

%\linenumbers

%_______________
%_______________
\section{Introduction}
The study of finite many-body systems \cite{2004Ring} is a very hard task due to the huge number of correlations which have to be taken into account. The dimension of the many-body model space grows greatly with the dimension of the single particle model space and the number of particles \cite{1980Lawson}. Even when large scale shell model formalism is able to deal with large matrix diagonalization  \cite{2005Caurier}, the study of a variety of nuclear properties of heavy nuclei is still prohibitive. 

In order to keep under control the dimension one may appeal to consider a simplified interaction, like the pairing force. It was immediately recognized from the seminal papers of Cooper \cite{1956Cooper} and Bardeen, Cooper and Schrieffer (BCS) \cite{1957Bardeen-Letter}, that the pairing concept could also be applied to nuclei \cite{1958Bohr}. Many nuclear properties were explained in term of a generalized Cooper's pair concept in finite nuclei \cite{1959Belyaev,1964Lane}. Besides, the pairing is one of the main contribution to the residual interaction \cite{1960Kisslinger,1963Kisslinger} and it is still nowadays, an important ingredient for the study of finite nuclei and nuclear matter \cite{2003Dean,2005Brink,2019Muther}.
  
Different approaches have been developed in the course of the last fifty years to deal with the pairing for finite \cite{2013Zelevinsky} and infinite systems \cite{2010Cooper}. The algebraic field theory approach, developed by Gorkov \cite{1958Gorkov,1971Fetter,2011Gorkov}, is one of the less exploited in finite systems and the one to be used in this work.

An alternative to the algebraic approach \cite{1958Gorkov,1966Irvine} to the Gorkov treatment of the pairing, is the perturbative one. This approach is an extension of the self-consistent Green's function method \cite{2004Dickhoff,2005Dickhoff} and it has been recently developed up to second order by Soma and collaborators in a series of papers \cite{2011Soma,2013Soma,2014Soma} . At first order, the perturbative version of the Gorkov equations coincides with the Hartree-Fock-Bogoliubov one \cite{2011Soma,1999Bulgac,2013Dobaczewski}. 

In this work we adopt the original quantum field theory approach of Gorkov \cite{1958Gorkov} with some modifications and adapted to spherical finite system. Our approach is closely related to the one given in Ref. \cite{1966Irvine}. Previous uses of the anomalous propagator in superfluid nuclear matter may be found in Refs. \cite{1999Bozek,2002Bozek}. The goal of this paper is to get algebraic equations for the separable pairing interaction \cite{1998Pan,2007BalantekinJesus,2019Pan} in finite system, as the first step for developing an extension, which take into account the continuum spectrum of energy, similar to Ref.  \cite{1987Belyaev} but in the complex energy representation \cite{1968Berggren}. The exactly solvable degenerate and symmetric \cite{1965Lipkin} models will be used as non trivial test systems to illustrate our approach.

In Section \ref{sec.formalism} we get the algebraic Gorkov solution for a spherical finite system with separable pairing interaction. In Sec. \ref{sec.observable} we show how to calculate different observables within this approach. The applications to the degenerate and the two levels models are developed in Sec. \ref{sec.application}. In the last Section (Section \ref{sec.conclusion})  some conclusions are drawn, and we sketch some perspectives for the future. A few appendices are included to show how to obtain some expressions given in the main body of the paper.

%_____________
%_____________
\section{Formalism} \label{sec.formalism}
The natural line of action in a particle conserving many-body system, described by the Hamiltonian $H$, would be to seek a many-body wave function $|\Psi_N\rangle$ which satisfies the associate conservation law, i.e. $\hat{N} |\Psi_N\rangle = N |\Psi_N\rangle$. But, in the BCS we have an example where a wave function which violate the conservation law does a good job, even in finite systems. This paradox can be understood in the framework of Lipkin approach to the study of collective motion in many-body systems \cite{1960Lipkin}.

\subsection{Ground state energy} \label{sec.formalism1}
Within the Lipkin approach \cite{1960Lipkin}, a model wave function $| \Psi \rangle$, which violates the conserving particle number, is defined; together with a model Hamiltonian $\mathcal{H}$. The modified wave function and the modified Hamiltonian allow the calculation of the properties of the original system described by the Hamiltonian $H$. Then, magnitudes calculated with $| \Psi \rangle$, like for example, $\langle \Psi | \hat{N} | \Psi \rangle=N$, are interpreted as averages over a set of neighboring nuclei.

Let us define our model Hamiltonian as $\mathcal{H}=H-f(\hat{N})$, with $f(\hat{N}) |\Psi_N\rangle = f(N) |\Psi_N\rangle$, and let us expand 
$| \Psi \rangle = \sum_N c_N  | \Psi_N \rangle$, with $\sum_N |c_N|^2=1$. Then, the following identity is verified 
\begin{equation}
   \langle \Psi | \mathcal{H} | \Psi \rangle=\langle \Psi_N | H-f(\hat{N}) | \Psi_N \rangle
\end{equation} 

Under the assumption that $\mathcal{E}=\langle \Psi | \mathcal{H} | \Psi \rangle$ is easier to calculate than $E=\langle \Psi_N | H | \Psi_N \rangle$, we will get the magnitude of interest as
\begin{equation}
    E = \mathcal{E} + f(N)
\end{equation}

The drawback of this approach is that the function $f(\hat{N})$ in not known and then we have to appeal to Taylor expansion \cite{1960Lipkin}. For the linear approximation we get, $\mathcal{H}=H-\lambda \hat{N}$, with $\lambda$ be fixed by the constraint $\langle \Psi | \hat{N} | \Psi \rangle=N$. To second order expansion of $f(\hat{N})$ we get the Lipkin-Nogami (LN) formulation \cite{1964Nogami} of the pairing. In Ref. \cite{2017IdBetan} it is shown, for the Lipkin model \cite{1965Lipkin}, that the LN approximation is very similar to the exact Richardson solution \cite{1963Vol3Richardson}.

For the development of the present formulation we will assume $\mathcal{H}=H-\lambda \hat{N}$ for the model Hamiltonian, and for the model wave function $| \Psi \rangle$ we will consider a non-conserving particle number ground state in the Fock space. 

A benefit of the Gorkov formulation of the pairing is that there is no need to give and explicit ansatz for $| \Psi \rangle$ as in BCS \cite{1957Bardeen}, neither it is necessary to define explicitly a quasiparticle for which $| \Psi \rangle$ is the vacuum, like in the Bogoliubov \cite{1958Bogoliubov} framework. Instead, two Green functions will be defined, in terms of which the equation of motion is written. By usual Fourier transformation we will change to an algebraic system of equations in order to obtain explicitly the Green function and $\lambda$. Finally, using Green function technique \cite{2005Dickhoff}, we will calculate the modified energy $\mathcal{E}=\langle \Psi | \mathcal{H} | \Psi \rangle$ and from it, and the above calculated $\lambda$, we will get the ground state energy of the system with $N$ particles as,
\begin{equation}
   E = \mathcal{E} + \lambda N
\end{equation}

%________________
\subsection{Interaction}\label{sec.formalism2}
The many-body Hamiltonian model for a spherical system with only two-body interaction, in second quantification notation, reads
\begin{align}
   \mathcal{H} &= H_0 + V - \lambda \hat{N} =  \mathcal{H}_0 + V  \\
       &= \sum_\alpha e_a 
       					c^\dagger_\alpha c_\alpha
       			+ \frac{1}{4} \sum_{\alpha \beta \gamma \delta}
       			   \bar{V}_{\alpha \beta \gamma \delta}
       			    c^\dagger_\alpha c^\dagger_\beta
       			    c_\delta c_\gamma		
\end{align}
with $\mathcal{H}_0=H_0- \lambda \hat{N}$, $e_a=\varepsilon_a - \lambda$, $\alpha=\{ a, m_\alpha \}= \{ n_a,l_a,j_a,m_\alpha \}$, and $\bar{V}_{\alpha \beta \gamma \delta}=\langle \alpha \beta |V| \gamma \delta \rangle$ the \textit{antisymmetric} matrix elements of the two-body interaction and $|\alpha \beta\rangle=c^\dagger_\alpha c^\dagger_\beta | 0 \rangle$. The single particle operator $c^\dagger_\alpha$ creates a particle with wave function 
$\varphi_\alpha(\bs{r})=\frac{u_a(r)}{r} [Y_{l_a}(\hat{r}) \chi_{1/2}]_{j_a m_\alpha}$ which is eigenfunction of the single-particle mean-field Hamiltonian $h(\bs{r})$ of $H_0=\sum_i h(\bs{r}_i)$, i.e. $h(\bs{r}) \varphi_\alpha(\bs{r}) =  \varepsilon_a \varphi_\alpha(\bs{r})$.

By writing the generalized \textit{antisymmetric} pairing interaction matrix elements \cite{1964Lane},  $\langle ab,J | V | cd, J \rangle = \frac{g_{ac}}{2} \delta_{J0} \hat{j}_a \hat{j}_c$ in terms of the uncoupled \textit{antisymmetric} matrix elements \cite{2007Suhonen}, we get
\begin{equation} \label{eq.sf1}
    \bar{V}_{\alpha \beta \gamma \delta} = g_{ac} 
          \delta_{\alpha \bar{\beta}}
          \delta_{\gamma \bar{\delta}}
\end{equation}
with $m_\alpha>0$, and $m_\gamma>0$, since $\bar{V}$ is asymmetric, i.e. $\bar{V}_{\alpha \beta \gamma \delta}=-\bar{V}_{\beta \alpha \gamma \delta}=-\bar{V}_{\alpha \beta \delta \gamma }$. The functional $\delta_{\alpha \bar{\beta}}$ is a short hand notation for 
\begin{equation}\label{eq.sf2}
  \delta_{\alpha \bar{\beta}} 
  				= (-)^{j_a-m_\alpha} \delta_{a b} \delta_{m_\alpha, -m_\beta}
\end{equation}

For the separable pairing interaction the state dependent strength reads \cite{1998Pan,2007BalantekinJesus,2019Pan},
\begin{equation}\label{eq.sf3}
   g_{ac} = -g f_a f_c
\end{equation}
with  $g>0$ and $f_a = \int dr \, u^2_a(r) f(r)>0$, where the radial form factor function $f(r)$ is taken, for simplicity, to be positive defined. Different shapes of $f(r)$ can accommodate volume or surface pairing \cite{1987Belyaev}.

Using Eqs. (\ref{eq.sf1})-(\ref{eq.sf3}), we recover the usual pairing Hamiltonian,
\begin{align} \label{eq.sf4}
   \mathcal{H} &=\sum_\alpha e_a 
       					c^\dagger_\alpha c_\alpha
       			-  g \, P^\dagger \, P
\end{align}
with $P=(P^\dagger)^\dagger$ and
\begin{align}
   P^\dagger =& \sum_{a,m_\alpha>0} f_a \, c^\dagger_\alpha c^\dagger_{\bar{\alpha}}
\end{align}
where $c^\dagger_{\bar{\alpha}}$ is the time reverse state of $c^\dagger_{\alpha}$, i.e. $c^\dagger_{\bar{\alpha}}=(-)^{j_a-m_\alpha} c^\dagger_{a, -m_\alpha}$.

Equation (\ref{eq.sf4}) gives support to the matrix elements of the separable interaction (\ref{eq.sf1}) to be used in the next subsection to build the equation of motion.

%____________________
\subsection{Green functions}
The formalism is based on the many-body Green function technique. For this aim, we start by defining the usual one-particle Green function $G_{\alpha \alpha'}(t,t')$ and the two-particle correlation function or anomalous Green function $\tilde{G}^\dagger_{\alpha \alpha'}(t,t')$,
\begin{align}
   i\hbar \, G_{\alpha \alpha'}(t,t') &=
   		\langle \Psi | T  c_\alpha(t) c^\dagger_{\alpha'}(t')
   						 | \Psi \rangle \label{eq.g0}  \\
   %%%%
   	 i\hbar \,  \tilde{G}^\dagger_{\alpha \alpha'}(t,t') &=
   		\langle \Psi | T  c^\dagger_\alpha(t) c^\dagger_{\alpha'}(t')
   					 | \Psi \rangle	\label{eq.f}				 
\end{align}
The anomalous Green function is defined with the same phase as for the Green function just to keep the definition symmetric. In Refs. \cite{1966Irvine} and \cite{1987Belyaev}, for example, the phase is taken real. Another difference, is that the correlated ground state is not eigenfunction of the particle number operator, $\hat{N} | \Psi \rangle \ne N  | \Psi \rangle$. Besides, the anomalous Green function is not defined between time reversed states, as for example in Ref. \cite{2011Soma} but, we will find that its antisymmetric character will appear as a consequence of the pairing interaction, as in the original work of Gorkov \cite{1958Gorkov}.

The time ordering operator $T$ in Eqs. (\ref{eq.g0})  and (\ref{eq.f}) is defined as usual,
\begin{equation} \label{eq.t0}
   T  c_\alpha(t) c^\dagger_{\alpha'}(t') =
        \theta(t-t') c_\alpha(t) c^\dagger_{\alpha'}(t')
        - \theta(t'-t) c^\dagger_{\alpha'}(t')  c_\alpha(t) 
\end{equation}
with $c^\dagger_{\alpha}(t)$ the Heisenberg representation of the creation operator evolved with the model Hamiltonian $\mathcal{H}$,
\begin{equation}
     c^\dagger_{\alpha}(t) = e^{\frac{i}{\hbar}\mathcal{H} t}
     		c^\dagger_{\alpha} e^{-\frac{i}{\hbar}\mathcal{H} t}
\end{equation}

%______________________
\subsection{Equation of motion} \label{sec.eqmotion}
In this subsection we will derive the equations of motion for $G_{\alpha \alpha'}(t,t')$ and $\tilde{G}^\dagger_{\alpha \alpha'}(t,t')$ for the separable pairing interaction (\ref{eq.sf1})-(\ref{eq.sf3})
\begin{equation} \label{eq.sf4b}
    \bar{V}_{\alpha \beta \gamma \delta} =  -g  f_a  f_c
          (-)^{j_a-m_\alpha} (-)^{j_c-m_\gamma}
          \delta_{a b} \delta_{c d}
          \delta_{m_\alpha, -m_\beta} \delta_{m_\gamma, -m_\delta}
          \hspace{8mm}
          m_\alpha>0
          \hspace{3mm}
          m_\gamma>0
\end{equation}

By using explicitly the time ordering (\ref{eq.t0}) and taking the time derivative of Eqs. (\ref{eq.g0}) and (\ref{eq.f}) with respect to $t$, and the identities shown in Appendix \ref{app.1}, we get the following two coupled-equation for $G$ and $\tilde{G}$,
\begin{align}
    i\hbar \,\frac{\partial }{\partial t}G_{\alpha \alpha'}(t,t') &=
    		 \delta(t-t') \, \delta_{\alpha \alpha'}
            + e_a G_{\alpha \alpha'}(t,t')   \nonumber \\
      &+ \sum_{\beta \gamma \delta} \, 
                 \frac{\bar{V}_{\alpha \beta \gamma \delta}}{2i \hbar} 
	 \langle \Psi | 
	 T [ c^\dagger_\beta(t) c_\delta(t) c_\gamma(t) c^\dagger_{\alpha'}(t')]  | \Psi \rangle
	 \label{eq.dgdt} \\
	 %%%
	 i\hbar \,\frac{\partial }{\partial t} \tilde{G}^\dagger_{\alpha \alpha'}(t,t')  &=
	 	- e_a \, \tilde{G}^\dagger_{\alpha \alpha'}(t,t')  \nonumber \\
        & - \sum_{\beta \gamma \delta} \, 
                \frac{\bar{V}_{\alpha \beta \gamma \delta}}{2i \hbar} 
   \langle \Psi | 
	 T [ c^\dagger_\gamma(t) c^\dagger_\delta(t) 
        	 c_\beta(t) c^\dagger_{\alpha'}(t')]
 | \Psi \rangle \label{eq.dfdt}        
\end{align}	 

Next, we evaluate these equations for the separable pairing interaction $\bar{V}_{\alpha \beta \gamma \delta}$ as given by Eq. (\ref{eq.sf4b}),
\begin{align}
   i\hbar \,\frac{\partial }{\partial t} G_{\alpha \alpha'}(t,t') &=
           \delta(t-t') \, \delta_{\alpha \alpha'}
            + e_a G_{\alpha \alpha'}(t,t') \nonumber \\
     & - \frac{g\, f_a}{i \hbar}
       \sum_{\gamma>0} \, f_c\,
    	 \langle \Psi |  T [ c^\dagger_{\bar{\alpha}}(t) c_{\bar{\gamma}}(t) 
	 		c_{\gamma}(t) c^\dagger_{\alpha'}(t')] | \Psi \rangle \label{eq.dgdta}  \\
	%%%%%
   i\hbar \,\frac{\partial }{\partial t} \tilde{G}^\dagger_{\alpha \alpha'}(t,t') &=
            - e_a \tilde{G}^\dagger_{\alpha \alpha'}(t,t') \nonumber \\
     &  + \frac{g\, f_a}{i \hbar} 
       \sum_{\gamma>0} \, f_c\,
    	 \langle \Psi |  T [ c^\dagger_{\gamma}(t) c^\dagger_{\bar{\gamma}}(t) 
	 		c_{\bar{\alpha}}(t) c^\dagger_{\alpha'}(t')] | \Psi \rangle \label{eq.dfdta}
\end{align}

In the mean-field approximation \cite{1958Gorkov,1971Fetter}, the above equations of motion can be written fully in terms of $G$ and $\tilde{G}$ (see appendix \ref{app.2}.)

\begin{align}
  & \left( i\hbar \frac{\partial }{\partial \tau}  - \tilde{e}_a  \right) G_{\alpha \alpha'}(\tau) 
      - \frac{\hbar g f_a}{i} 
         \sum_{\gamma>0} \, f_c \,
                \tilde{G}_{\bar{\gamma} \gamma}(0)
                \tilde{G}^\dagger_{\bar{\alpha} \alpha'}(\tau)  =
           \delta(\tau) \delta_{\alpha \alpha'}   \label{eq.dgdt5} \\
   %%%%
   \nonumber \\
   %%%%
  &  \left(  i\hbar \frac{\partial }{\partial \tau} + \tilde{e}_a \right) 
   				\tilde{G}^\dagger_{\alpha \alpha'}(\tau)  
  + \frac{\hbar g f_a}{i}  
      \sum_{\gamma>0} \, f_c \,
           \tilde{G}^\dagger_{\gamma \bar{\gamma}}(0) 
           				G_{\bar{\alpha} \alpha'}(\tau)   =0      \label{eq.dfdt5}
\end{align}
with $m_\alpha>0$, $\tau=t-t'$ and $\tilde{e}_a=e_a - g f^2_a  \, n_a$, the modified single particle energy, with $n_a=-i\hbar G_{\alpha \alpha}(0^-)$ the occupation probability of the state $\alpha$. The magnitude $\tilde{G}^\dagger_{\gamma \bar{\gamma}}(0)$ has to be understood as the following limit,
\begin{equation}
   \tilde{G}^\dagger_{\gamma \bar{\gamma}}(0) 
   			= lim_{\tau \rightarrow 0} \tilde{G}^\dagger_{\gamma \bar{\gamma}}(\tau) \, , 
\end{equation}
later we will show that $\tilde{G}^\dagger_{\gamma \bar{\gamma}}(0)$ does not depend on how we approach zero, .i.e., 
$ \tilde{G}^\dagger_{\gamma \bar{\gamma}}(0^+) 
= \tilde{G}^\dagger_{\gamma \bar{\gamma}}(0^-)$. 

The above equations (\ref{eq.dgdt5}) and (\ref{eq.dfdt5}) constitute the final version of the Gorkov equations of motion in time dependent representation. Next we will solve them by performing the Fourier transform.

 %______________________________
 \subsection{Algebraic solution of the equation of motion}
Let us introduce the state dependent gap parameters $\Delta_a$,
\begin{equation}
		 \Delta_a = \frac{f_a \, g \, \hbar}{i} \, \sum_{\gamma>0} \, f_c \,
              \tilde{G}_{\bar{\gamma} \gamma}(0) 
\end{equation}
Using the relations 
$\left( \tilde{G}_{\bar{\gamma} \gamma}(0) \right)^*
=\tilde{G}^\dagger_{\bar{\gamma} \gamma}(0)
=-\tilde{G}^\dagger_{ \gamma \bar{\gamma}}(0)$ we can write its complex conjugate
\begin{equation}
    \Delta^*_a= \frac{f_a \, g \, \hbar}{i}\, \sum_{\gamma>0} \, f_c \,
              \tilde{G}^\dagger_{\gamma \bar{\gamma}}(0) \label{eq.d}
\end{equation}

With the above definitions, the equations of motion (\ref{eq.dgdt5}) and (\ref{eq.dfdt5}) read,
\begin{align}
   \left( i\hbar \frac{\partial }{\partial \tau}  - \tilde{e}_a  \right) G_{\alpha \alpha'}(\tau) 
       - \Delta_a \,  \tilde{G}^\dagger_{\bar{\alpha} \alpha'}(\tau) &=
                   \delta(\tau) \, \delta_{\alpha \alpha'} \hspace{6mm} \label{eq.dgdt6} \\
     %%%%%
     \left(  i\hbar \frac{\partial }{\partial \tau} + \tilde{e}_a  \right)
          \tilde{G}^\dagger_{\alpha \alpha'}(\tau)
      + \Delta^*_a  \,  G_{\bar{\alpha} \alpha'}(\tau)   &= 0 \label{eq.dfdt6}
\end{align}

Next, we make the Fourier transform to energy
\begin{equation}
    G_{\alpha \alpha'}(\tau) = 
        \int_{-\infty}^\infty   G_{\alpha \alpha'}(E) \, e^{-i \frac{E}{\hbar} \tau} \, 
             \frac{dE}{2\pi\hbar}
\end{equation}
of the equations of motion (\ref{eq.dgdt6}) and (\ref{eq.dfdt6}) and get,
\begin{align}
   (E - \tilde{e}_a) G_{\alpha \alpha'}(E) 
   				- \Delta_a \tilde{G}^\dagger_{\bar{\alpha} \alpha'}(E) 
   							&= \delta_{\alpha \alpha'}  \label{eq.dgdt7}   \\
   %%%%%
   (E + \tilde{e}_a ) \tilde{G}^\dagger_{\alpha \alpha'}(E)
        +  \Delta^*_a  G_{\bar{\alpha} \alpha'}(E) &= 0	\label{eq.dfdt7}
\end{align}
where we have used $\delta(\tau)=\int \exp(-i\frac{E}{\hbar}\tau) \frac{dE}{2\pi \hbar}$.

Since 
$G_{\bar{\alpha} \alpha'}(E)=\delta_{\bar{\alpha} \alpha'} G_{\bar{\alpha} \bar{\alpha}}(E)$, Eq. (\ref{eq.dfdt7}) gives for $\bar{\alpha} \ne \alpha' \Rightarrow \tilde{G}^\dagger_{\alpha \alpha'}(E)=0$, which shows that the only non-zero matrix elements of $\tilde{G}^\dagger$ are time reversed companion states, then 
\begin{equation}
		\tilde{G}^\dagger_{\alpha \alpha'}(E) = 
   			\delta_{\bar{\alpha} \alpha'} \tilde{G}^\dagger_{\alpha \bar{\alpha}}(E)  \label{eq.hoy}
\end{equation}
 
Equation (\ref{eq.dgdt7}), for $\alpha \ne \alpha'$ gives 
$\tilde{G}^\dagger_{\bar{\alpha} \alpha'}(E)=0$, showing that the only non-zero matrix elements of $\tilde{G}^\dagger$ are time reversed companion states. Then 
$ \tilde{G}^\dagger_{\bar{\alpha} \alpha'}(E) = 
   			\delta_{\alpha \alpha'} \tilde{G}^\dagger_{\bar{\alpha} \alpha}(E)$, which is like Eq. (\ref{eq.hoy}) by the substitution $\bar{\alpha} \leftrightarrow \alpha$.

Using the above identity we get for the equations of motion,
\begin{align}
   (E - \tilde{e}_a) G_{\alpha \alpha}(E) 
   				+ \Delta_a \tilde{G}^\dagger_{ \alpha \bar{\alpha}}(E) 
   							&= 1  \label{eq.dgdt8}   \\
   %%%%%
   (E + \tilde{e}_a ) \tilde{G}^\dagger_{\alpha \bar{\alpha}}(E)
        +  \Delta^*_a  G_{\alpha \alpha}(E) &= 0	\label{eq.dfdt8}
\end{align}
where we have used $\tilde{G}^\dagger_{\bar{\alpha} \alpha }(E)=-\tilde{G}^\dagger_{\alpha \bar{\alpha}}(E)$ and $G_{\bar{\alpha} \bar{\alpha}}(E)=G_{\alpha \alpha}(E)$. 

These are two algebraic coupled equations for $G_{\alpha \alpha}(E) $ and $\tilde{G}^\dagger_{\alpha \bar{\alpha}}(E)$ for $m_\alpha>0$, which written in matrix form, resembles very much that of Ref. \cite{1987Belyaev} but in energy representation. The solution of the above system gives,
\begin{align}
   G_{\alpha \alpha}(E) &= \frac{E+\tilde{e}_a}{E^2-E^2_a} \label{eq.gz}  \\
   %%%
   \tilde{G}^\dagger_{\alpha \bar{\alpha}}(E) 
   				&= - \frac{\Delta^*_a}{E^2-E^2_a}
\end{align}
where we have introduced the shifted (quasiparticle) energy $E^2_a=\tilde{e}^2_a+ |\Delta_a|^2$. 

By making a partial fraction decomposition of Eq. (\ref{eq.gz}), 
\begin{equation}
    G_{\alpha \alpha}(E) = \frac{A_a}{E-E_a}
    				+ \frac{B_a}{E+E_a}
\end{equation}
we get
\begin{align} 
    A_a &= \frac{E_a+\tilde{e}_a}{2E_a} \\
    B _a&= \frac{E_a-\tilde{e}_a}{2E_a} \label{eq.b} 
\end{align}
with $A_a+B_a=1$. 

At this stage we can recognize the BCS occupation probability
\begin{equation}
     B_a=\frac{E_a-\tilde{e}_a}{2E_a}
             = \frac{1}{2} \left(  1 - \frac{\tilde{e}_a}{E_a} \right)
             =v^2_a \, ,
\end{equation}
and from $A_a+B_a=1 \Rightarrow u^2_a=A_a$, then
\begin{equation}
    G_{\alpha \alpha}(E) = \frac{u^2_a}{E-E_a}
    				+ \frac{v^2_a}{E+E_a}
\end{equation}

The above expression shows that the Green function has a pole at $E=E_a$ and at $E=-E_a$, where we have adopted the convention $E_a>0$. In order to perform the inverse Fourier transformation, we have to know how to avoid these singularities on the real $E$ axis where the two poles lie. For this purpose we extend the real variable $E$ to the complex plane, i.e. 
\begin{equation}
     E+i |\eta| \, \tr{sgn}(\eta) \, ,
\end{equation}
where $\eta$ is an infinitesimal. Its sign is determined by the Landau \cite{1958Landau} prescription 
\begin{equation}
      \frac{\Im[G(E)]}{E} < 0  \, .
\end{equation}

Using the identity $(x-x_0+i 0)^{-1}=P\frac{1}{x-x_0}-i \pi \delta(x-x_0)$ we get,
\begin{align}
    \frac{\Im[G(E)]}{E} &= -\pi \frac{\tr{sgn}(\eta)}{E}
      \left[
          u^2_a \delta(E-E_a) \right. 
           \left. + v^2_a \delta(E+E_a) 
      \right] \nonumber \\
       \Rightarrow  \tr{sgn}(\eta)&= \tr{sgn}(E) \, .
\end{align}

Then, the Green function reads,
\begin{equation} 
    G_{\alpha \alpha}(E) = \frac{u^2_a}{E+i |\eta| \, \tr{sgn}(E)-E_a}
    			+ \frac{v^2_a}{E+i |\eta| \,\tr{sgn}(E)+E_a} \label{Eq.g}
\end{equation}

Due to the fact that $E_a>0$, the Green function has a pole at $E=E_a-i |\eta| \, \tr{sgn}(E)=E_a-i |\eta|$ with residues $u^2_a$ and a second pole at $E=-E_a-i |\eta| \, \tr{sgn}(E)=-E_a+i |\eta|$ with residues $v^2_a$. This result allow us to write the Green function in a more amenable and practical way,
\begin{equation} \label{Eq.g2}
    G_{\alpha \alpha}(E) = 
    				\frac{u^2_a}{E - (E_a - i \eta)}
    				+ \frac{v^2_a}{E - (-E_a + i \eta)}
\end{equation}
where now, $\eta$ is a positive infinitesimal.

The anomalous propagator reads,
\begin{align} \label{Eq.f2}
    \tilde{G}^\dagger_{\bar{\alpha} \alpha}(E) &=
    				- \frac{\Delta^*_a}
    				{[E - (E_a - i \eta)][E - (-E_a + i \eta)]}
\end{align}
which shows that $\tilde{G}^\dagger_{\bar{\alpha} \alpha}(E)$ has a pole at $E=E_a - i \eta$ with residues $-\frac{\Delta^*_a}{2E_a }$ and another pole at $E=-E_a + i \eta$, with residues $\frac{\Delta^*_a}{2E_a}$ (we have taken the $\lim_{\eta \rightarrow 0}$).

Equations (\ref{Eq.g2}) and (\ref{Eq.f2}) fully solved the algebraic equations of motion (\ref{eq.dgdt8}) and (\ref{eq.dfdt8}).

 %______________
 %______________
\section{Observables} \label{sec.observable}
In this section we will build up the explicit equations which allow to solve the many-body problem for a given number of particles and interaction $g_{ab}=-gf_af_b$. The ground state energy will be written as a close equation in terms of the quasiparticle energies $E_a$, Fermi level $\lambda$ and particle number $N$.

%___________________________
\subsection{Particle number equation}
The non-conserving particle number ground-state $| \Psi \rangle$ must gives, in average the right particle number $N$, i.e. $\langle \Psi | \hat{N} | \Psi \rangle = N$, then
\begin{equation} \label{eq.NN}
   N  =  \sum_\alpha n_\alpha = -i \, \hbar \, \sum_\alpha G_{\alpha \alpha}(\tau=0^-) 
\end{equation}

By integrating $G_{\alpha \alpha}(\tau)$ in a closed contour in the upper complex $E$ plane, because it is the region for which the exponential $e^{-i\frac{E}{\hbar}\tau}=e^{i\frac{E}{\hbar} |\tau|}$ converges, we get
\begin{align}
    G_{\alpha \alpha}(0^-) &= \lim_{\tau \rightarrow 0^-} 
    					G_{\alpha \alpha}(\tau) \nonumber \\
      %%%%
      &= \lim_{\tau \rightarrow 0^-}
             \int%_{-\infty}^\infty
             \left[
                \frac{u^2_a}{E - (E_a - i |\eta|)}
    				+ \frac{v^2_a}{E - (-E_a + i |\eta|)}
             \right] 
              e^{-i\frac{E}{\hbar}\tau} \frac{dE}{2\pi \hbar} \nonumber \\
    %%%%
    &= \frac{i}{\hbar} v^2_a      \label{eq.v2}
\end{align}
then, $n_\alpha= v^2_a$. 

By replacing (\ref{eq.v2}) in Eq. (\ref{eq.NN}), we get for the particle number average,
\begin{equation}
   N = \sum_\alpha v^2_a \label{eq.n}
\end{equation}
This equation fixes the Fermi level $\lambda$ for the given number of particles of the system in terms of the known parameters $\Delta_a$ (which are implicit in $v^2_a$). The next step is to find out a set of equations for the parameters $\Delta_a$.

%_____________________
\subsection{Gap equation} \label{sec.gap}
We will proceed as Gorkov did in Ref. \cite{1958Gorkov} in order to get another independent equation to solve the many-body problem, which consist into using the state dependent gap equation (\ref{eq.d}),
\begin{equation}
    \Delta^*_a = \frac{f_a \, g \, \hbar}{i}\, \sum_{\gamma>0} \, f_c \,
              \tilde{G}^\dagger_{\gamma \bar{\gamma}}(\tau=0) 
\end{equation}

In the Appendix \ref{app.3} we show that $\tilde{G}^\dagger_{\gamma \bar{\gamma}}(0)$ is independent of how the limit is performed, and we found the value
\begin{equation}\label{eq.g0b}
    \tilde{G}^\dagger_{\gamma \bar{\gamma}}(0) 
        =  \frac{i}{ \hbar} \frac{\Delta^*_c}{2E_c} 
\end{equation}
then
\begin{equation}
    \Delta^*_a = f_a \, g \, \sum_{\gamma>0} \, f_c \,
              \frac{\Delta^*_c}{2E_c} 
\end{equation}

If we write,
\begin{equation}
  \Delta_a = f_a D \label{eq.gap3}
\end{equation}
with $D$ unknown, we get the gap equation
\begin{equation} \label{eq.gap}
    \frac{2}{g} =  \sum_{\gamma>0} \, \frac{f^2_c}{E_c}           
\end{equation}
which gives (for $D \ne 0$) an equation for $|D|^2$, since this parameter $D$ is implicit in $E_c=\sqrt{\tilde{e}^2_c+f^2_c |D|^2}$.

At the end, we have found that there is not as many equations as degree of freedom but only two.  Equations (\ref{eq.n}) and (\ref{eq.gap}) are the only two equations which are needed for the calculation of the two unknown parameters $\lambda$ and $D$. Then, with the parameter $D$ we can calculate each state dependent gap through Eq. (\ref{eq.gap3}) and then $v^2_a$, $\tilde{e}_a$ and $E_a$.

\textit{The following set of equations solve the finite many-body problem for the separable pairing interaction (let us take $D \in \mathfrak{R}$):}
\begin{align}
     N &= \sum_\alpha v^2_a  \\
      \frac{2}{g} &=  \sum_{\alpha>0} \, \frac{f^2_a}{E_a} \\
        v^2_a &= \frac{1}{2} \left(  1 - \frac{\tilde{e}_a}{E_a} \right) \\
      E_a &= \sqrt{\tilde{e}^2_a + f^2_a \ D^2} \\
      \tilde{e}_a &= \varepsilon_a - \lambda - g f_a v^2_a \label{Eq.58} \\
      \Delta_a &= f_a D
\end{align}
where $N$, $g$ and $f_a$ are given.

Next, we calculate the ground state energy from the Green function formalism and write it in terms of the above calculated parameters.

%_______________________
\subsection{Ground state energy} \label{sec.energy}
The correlated ground state energy $E$ is obtained from the average model energy $\mathcal{E}$ and the Fermi level $\lambda$,
\begin{equation}
   E =  \mathcal{E}  + \lambda N
\end{equation}

In Appendix \ref{app.4} we show that $\mathcal{E}$ can be calculated from the model wave function $|\Psi \rangle$ and the model Hamiltonian $\mathcal{H}$ using Green function technique,
\begin{equation}
   \mathcal{E} = \frac{1}{2}  \langle \Psi | \mathcal{H}_0  | \Psi \rangle 
   					+  \frac{1}{2} \sum_\alpha I_\alpha 
\end{equation}
with
\begin{align}
   \langle \Psi | \mathcal{H}_0  | \Psi \rangle &= \sum_\alpha \, v^2_a e_a  \\
    \sum_\alpha I_\alpha &=
     \frac{1}{\pi} \sum_\alpha  \int_{-\infty}^\lambda \, dE' \, E' \, \Im G_{\alpha \alpha}(E')
     =  - \sum_\alpha \, E_a \,  v_a^2 
\end{align}

Then, the correlated ground state energy in the non-conserving particle number ground state reads,
\begin{align}
   E &= \mathcal{E} + \lambda N \\
   %%%%
   &=  \frac{1}{2} \sum_\alpha \, v^2_a (\varepsilon_a - \lambda)
           		  - \frac{1}{2} \sum_\alpha \, E_a \,  v_a^2  		
           		  + \lambda N	\\
    %%%%
  E &=  \frac{1}{2} \sum_\alpha \, v^2_a (\varepsilon_a - E_a)
           		  + \frac{1}{2} \lambda N	
\end{align}
where $v^2_a$, $E_a$ and $\lambda$ are determined from the pairing gap and particle number equations and $N$ is the number of particles of the system.

%______________ 
%______________ 
\section{Applications} \label{sec.application}
In this section we will apply the above method to the degenerate and two level models, both of them are standard models usually used to illustrate many-body methods \cite{1961Feldman,1965Meshkov,1973Pradhan,2007Suhonen}. We will show the difference with the BCS solution.

%______________________
\subsection{Degenerate model}
As the first example of the above formulation we solve the degenerate model and we will compare it with its exact solution. Let as design $\varepsilon_\alpha=\varepsilon$ and $\sum_\alpha 1=2 \Omega$, where $\Omega$ is the pair degeneracy. In this model the only effect of the separable interaction amplitudes $f_a=f$ is to rescale the strength $g$ to $g'=gf^2$, then we will take $f_a=1$.

From the particle number constraint $N=\sum_\alpha v^2_a$ we get the particle occupation amplitude
\begin{equation}
   v_a =\sqrt{\frac{N}{2\Omega}}
\end{equation}

While, from the gap equation $\frac{2}{g}=\sum_{\alpha>0} \frac{f^2_a}{E_a}$ we get the quasiparticle energy
\begin{equation}
   E_a = \frac{g \Omega}{2}
\end{equation}

With the above results and the relation $2v_a^2=1-\tilde{e}_a/E_a$ we get the modified single particle energy $\tilde{e}_a=\frac{g \Omega}{2}(1-N/\Omega)$ and with $\tilde{e}_a=\varepsilon - \lambda - g v_a^2$ we get the Fermi level
\begin{equation}
   \lambda = \varepsilon - \frac{g \Omega}{2} 
   					\left( 1 - \frac{N}{\Omega} + \frac{N}{\Omega^2} \right)
\end{equation}

Now, we are in condition to calculate the ground state energy for any value of $N$,
\begin{align}
     E &= \frac{1}{2} \sum_\alpha v^2_a ( \varepsilon_a - E_a) + \frac{1}{2} \lambda N \\
     &= \frac{N}{2}\varepsilon - \frac{N}{4} g \Omega + \frac{N}{2} \lambda \\
     &= N \varepsilon - g \frac{N}{4} \left( 2 \Omega - N + \frac{N}{\Omega} \right)
\end{align}

This is like the usual BCS result. The comparison with the exact result \cite{2007Suhonen}, $E_{exact} = 	- \frac{g N}{4}  \left( 2\Omega -N + 2 \right)$ (for $\varepsilon=0$), shows that the relative error goes as the reciprocal of the pair degeneracy
\begin{equation}
    \frac{ E_{exact}-E}{ E_{exact}}
         = \left( \Omega + \frac{1}{1 - \frac{N}{2 \Omega}} \right)^{-1} 
         \approx \frac{1}{\Omega}
\end{equation}

Finally, from the quasiparticle relation $E_a^2=\tilde{e}_a^2 + D^2$ and $\Delta = f D=D$ we get the following value for the gap,
\begin{equation}
   \Delta = \frac{g N}{2} \sqrt{\frac{2 \Omega}{N} -1}
\end{equation}
which coincides with the one calculated in the usual BCS framework.

%______________________________
\subsection{The two level at half filling model}
The second application is a system with two levels $\varepsilon_{u/d}=\pm \frac{\varepsilon}{2}$ (with $\varepsilon>0$), each one with the same degeneracy $2\Omega$. We will give the solution for the special situation when the number of particles is $N=2\Omega$ which holds an algebraic solution and we will compare it with the BCS solution for constant pairing case.

From the particle number equation constraint $N=\sum_\alpha v^2_a$ we get $1=v^2_u + v^2_d$, which implies $v^2_u=u^2_d$ and $v^2_d=u^2_u$. By using any of these relations we get $\frac{\tilde{e}_u}{E_u} = - \frac{\tilde{e}_d}{E_d}$. By taking the reciprocal and squaring, using $E_{u/d}^2=\tilde{e}_{u/d}^2 + f_{u/d}^2 D^2$, we get $\tilde{e}_u = - \frac{f_u}{f_d} \tilde{e}_d$, which also implies $E_u = \frac{f_u}{f_d} E_d$.

Then, from the gap equation, 
$\frac{2}{g}=\sum_{\alpha_u>0} \frac{f_u^2}{E_u}+ \sum_{\alpha_d>0} \frac{f_d^2}{E_d}$, 
we get the quasiparticle energies
\begin{align}\label{eq.75}
    E_{u/d} &= \frac{g \Omega }{2} f_{u/d} (f_u + f_d)
\end{align}

The modified single particle energies read,
\begin{align}\label{eq.76}
    \tilde{e}_{u/d} &= \pm \frac{ f_{u/d} \Omega}{2}
    			\left[ \frac{2\varepsilon-g(f_u-f_d)}{(f_u+f_d)\Omega-1}  \right]
\end{align}

The Fermi level can be obtained from the expression 
$\tilde{e}_{u/d}=\varepsilon_{u/d} - \lambda - g f_{u/d} v^2_{u/d}$,
\begin{equation}
     \lambda = - \frac{\varepsilon	}{2}  \frac{f_u-f_d}{f_u+f_d}
                  -g  \frac{f_u f_d}{f_u+f_d}
\end{equation}

The occupation probability read,
\begin{equation}
    v^2_{u/d} = \frac{1}{2}
    			\left(1 \mp  \frac{\tilde{e}_u}{E_u} \right)
\end{equation}
with
\begin{equation}
   \frac{\tilde{e}_u}{E_u} = \frac{2 \varepsilon - g (f_u-f_d)}
   										{g(f_u+f_d)[\Omega(f_u+f_d)-1]}
\end{equation}

Using the above result we get for the ground state energy $E=\mathcal{E}+\lambda N $ (with $N=2\Omega$),
\begin{equation} \label{Eq.80}
     E = - \frac{\Omega}{2}\varepsilon
      					\left(   \frac{\tilde{e}_u}{E_u}
      					      + \frac{f_u-f_d}{f_u+f_d} \right)
      					      - \frac{\Omega}{2}E_u  \frac{f_u+f_d}{f_u}
      					     -  g \Omega \frac{f_u f_d}{f_u+f_d}
      					      + \frac{\Omega}{2} \tilde{e}_u \frac{f_u-f_d}{f_u} 
\end{equation}

From $E^2_d= \tilde{e}^2_d + f^2_d D^2$, we get for the constant $D$, % which give the state dependent pairing gap $\Delta_{u/d}$,
\begin{align}
   D^2  %=\frac{E^2_d - \tilde{e}^2_d}{f^2_d} 
             =\frac{E^2_d}{f^2_d} \left( 1 - \frac{\tilde{e}^2_d}{E^2_d} \right) 
            %= \frac{E^2_d}{f^2_d}
   			 %\left( 1 - \frac{\tilde{e}_d}{E_d} \right) 
   			 %\left( 1 + \frac{\tilde{e}_d}{E_d} \right) 
   = 4 \frac{E^2_d}{f^2_d} v^2_d  u^2_d
   %= \Omega^2 g^2 (f_u+f_d)^2 v^2_d  u^2_d
\end{align}
which can be written as,
\begin{equation}
     D = g \Omega (f_u+f_d) v_d u_d
\end{equation}
Since $D$ is positive defined, it is required  $u_d>0$, which implies that the strength has to be greater than $g_c$, with
\begin{equation}\label{Eq.83}
   g_c = \frac{2 \varepsilon}{\Omega (f_u+f_d)^2 - 2 f_d}
\end{equation}

%______________________________
\subsubsection{Constant pairing}
In this section we are going to compare the solution for the constant pairing interaction with that from the BCS approach \cite{2007Suhonen}.

For the particular case of constant pairing we have $g_{ab}=-gf_af_b=-g$, i.e. $f_a=f_b=1$, with $a$ and $b$ any of the two possible configurations \textit{up} or {\it down}. Then, using Eq. (\ref{eq.75}) we found that both quasiparticle energies are the same,$E_u=E_d=g \Omega$, which is the same that one gets from the BCS approach.

A straightforward substitution of the constant pairing in Eq. (\ref{eq.76}) give for the modified single particle energies
\begin{equation}
      \tilde{e}_u=-\tilde{e}_d=\frac{\Omega \varepsilon}{2\Omega-1} \label{Eq.84}
\end{equation}

This result differs from the BCS approach
\begin{equation}
   \tilde{e}^{BCS}_u=\frac{2\Omega \varepsilon}{4\Omega-1} \label{Eq.85}
\end{equation}

Even when both results coincides for $\Omega \gg 1$, the  discrepancy between the two approaches deserve a thoroughly analysis, because it will have an impact on the many-body correlated energy below.

By comparing Eq. (\ref{Eq.58}) for the modified single particle energy from our formulation, with the equivalent of Ref. \cite{2007Suhonen}, i.e. Eqs. (13.45) and Eqs. (13.105), we have
\begin{align}
     \tilde{e}_u &= \frac{\varepsilon}{2} - \lambda 
     		- g\, v^2_u \label{Eq.86} \\
     \tilde{e}^{BCS}_u &= \frac{\varepsilon}{2} - \lambda 
     				- g\, \left( v^2_u + \frac{v^2_d}{2} \right)  \label{Eq.87}
\end{align}

The extra term in Eq. (\ref{Eq.87}) has it origin in the off diagonal term of Eq. (13.45) in Ref. \cite{2007Suhonen}. Then using the Eqs. $v^2_u+v^2_d=1$ and $v^2_u=(1-\tilde{e}_u/E)/2$, which are common to both formulations,  we get Eqs. (\ref{Eq.84}) and (\ref{Eq.85}).

Similarly, the pairing gap coincides at the limit $\Omega \gg 1$ ($\Delta^2=g^2\Omega^2-\varepsilon^2/4$), but they differ in form according to,
\begin{align}
	\Delta^2 &= g^2 \Omega^2 
    			- \left( \frac{\Omega \varepsilon }{2 \Omega -1} \right)^2 \\
	\Delta^2_{BCS} &=g^2 \Omega^2 
            - \left( \frac{2 \Omega \varepsilon}{4 \Omega -1} \right)^2
\end{align}

By direct substitution of $f_u=f_d=1$ in Eqs. (\ref{Eq.80})  and (\ref{Eq.83}) we get for the many-body ground state energy,
\begin{equation} \label{eq.g}
    E_{\rm{Gorkov}}= - \frac{\Omega \varepsilon^2}{2g(2\Omega -1)}
    		- g \Omega ( \Omega + \frac{1}{2} ) \, ,
\end{equation}
provided that $g > g_c = \varepsilon/(2\Omega-1)$, Eq. (\ref{Eq.83}).

From Ref. \cite{2007Suhonen}, the homologous magnitude in the BCS approach is given by Eq. (13.119),
\begin{equation}
   E_{\rm{BCS}} = - \frac{3}{4} \Omega g 
   			- \frac{1}{2g}\, \varepsilon\, \tilde{e}^{BCS}_u\, \frac{8\Omega-1}{4\Omega-1}
   						- \Omega^2 g
   						+ \frac{1}{g} \left( \frac{2\Omega \varepsilon}{4\Omega-1} \right)^2
\end{equation}
by replacing $\tilde{e}^{BCS}_u$ from Eq. (\ref{Eq.85}) we get,
\begin{equation} \label{eq.bcs}
   E_{\rm{BCS}} = -  \frac{\Omega \varepsilon^2 }{g(4\Omega-1)}
   				-  g \Omega  \left( \Omega + \frac{3}{4} \right) \, ,
\end{equation}
as long as $g > 2\varepsilon/(4\Omega-1)$.

In the limit $\Omega \gg 1$, both many-body correlated energies (\ref{eq.g}) and (\ref{eq.bcs}) coincide and give (for $g > \varepsilon/(2\Omega)$),
\begin{equation} 
   E= -  \frac{\varepsilon^2 }{4g}
   				-  g \Omega^2   
\end{equation}

\begin{figure}[h!t]
%\vspace{6mm}
\begin{center}
    \includegraphics[angle=-90,width=0.75\textwidth]{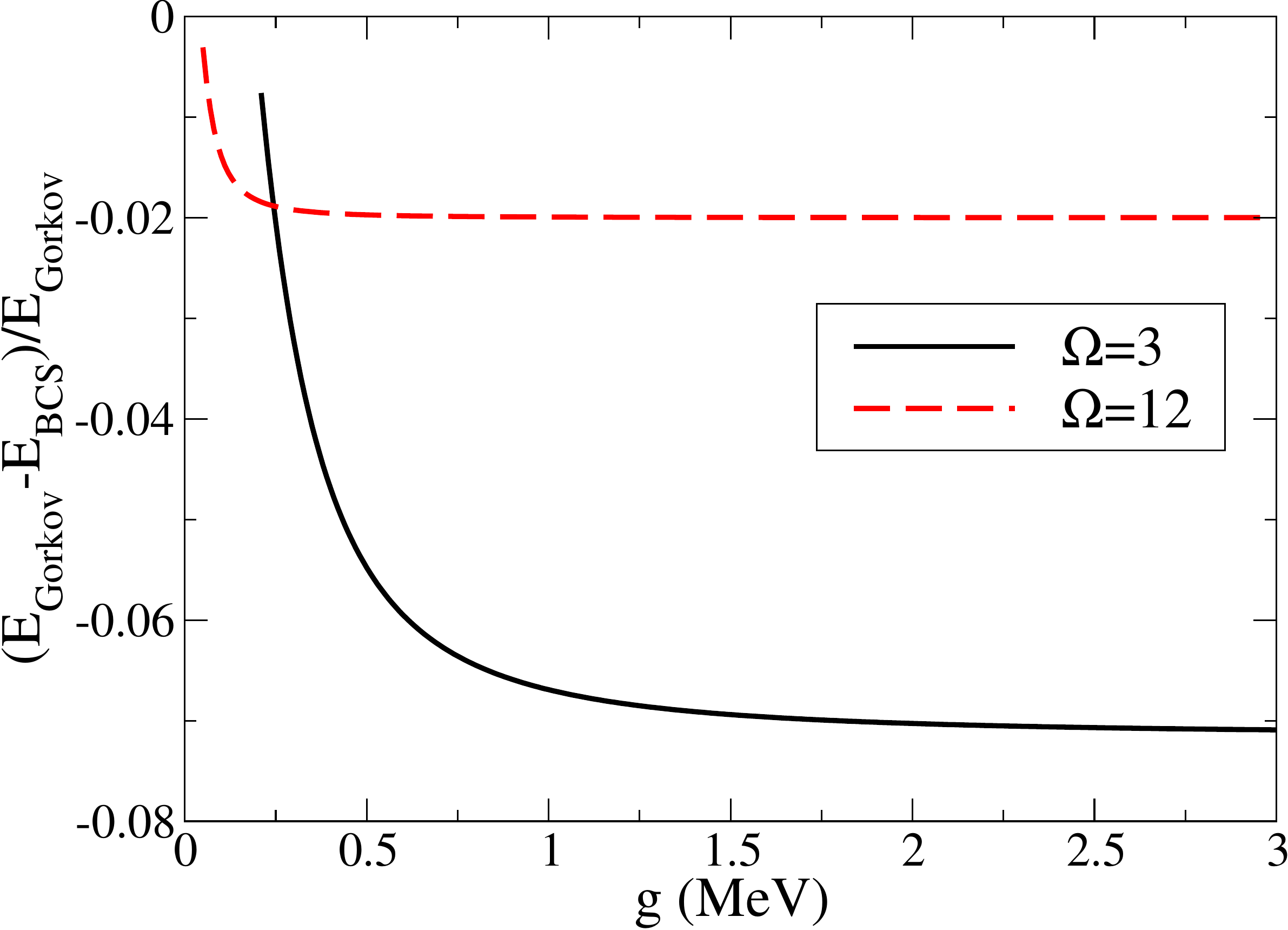}
\end{center}                
\caption{\label{fig.G-BCS} (Color online) Relative difference between Gorkov and BCS ground state energies for $\Omega = 3$ and $\Omega = 12$ as a function of the pairing strength $g$.}
%\vspace{6mm}
\end{figure}

In Fig. \ref{fig.G-BCS} we compare the energy from the Gorkov and BCS approaches for two different value of degeneracy. We found that the relative energy differ in less than 7\% for low degeneracy, $\Omega=3$, quickly approaching to a constant value. A similar result is found when we increase the degeneracy to $\Omega=12$, but now the difference is less than two percent and one approaches to the other for much smaller values of the strength.

\begin{figure}[h!t]
%\vspace{6mm}
\begin{center}
    \includegraphics[angle=0,width=0.75\textwidth]{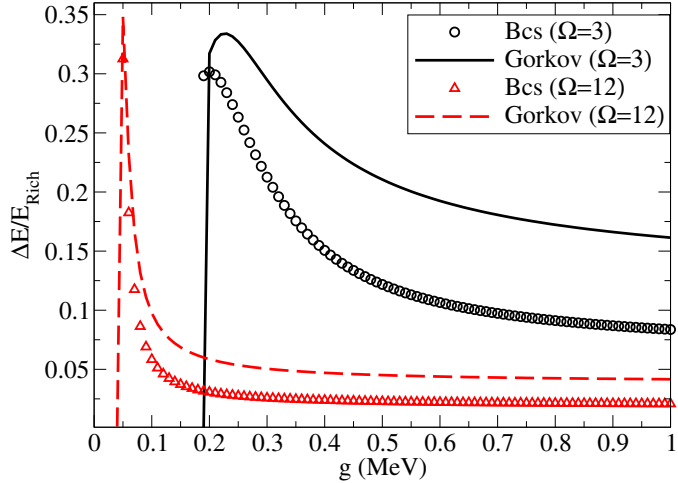}
\end{center}                
\caption{\label{fig.egs} (Color online) Relative Gorkov and BCS ground state energies with respect to the exact one, for $\Omega=3$ and $\Omega=12$ as a function of the pairing strength $g$.}
%\vspace{6mm}
\end{figure}

As a last analysis let us compare the Gorkov energy with the exact solution from the Richardson approach \cite{1963Vol3Richardson,2017IdBetan}. From Fig. \ref{fig.egs} we learn that the difference with respect to exact solution quickly drops from around 30 \% to 5\% as the degeneracy increases. A similar comparison with the BCS solution is also shown in Fig. \ref{fig.egs}. While the overall feature of the improvement with the degeneracy is the same, we found that the BCS approach does a better job than Gorkov.

From the sign of the relative energy shown in Fig. \ref{fig.G-BCS} we know that $|E_{\rm{BCS}}|>|E_{\rm{Gorkov}}|$, i.e. the BCS approach provides extra correlations which are missing in the Gorkov one. Tracing back the origin of the difference between the BCS and Gorkov solutions, we found that it is due to the fact that the constant pairing approach, as developed in Ref. \cite{2007Suhonen}, uses the interaction 
$\langle aa,0 | V| cc,0 \rangle = -g \Omega$, where $a,\, c$ are any of the \textit{upp} or \textit{down} levels, while in our Gorkov formulation we keep only the diagonal part of the interaction, i.e.
$\langle aa,0 | V| cc,0 \rangle = -g \Omega\ \delta_{ac}$. Then, the modified single particle energy (\ref{Eq.86})  in our formulation is corrected only for a single diagonal term $g f^2_a n_a$ (see Eq. (\ref{eq.99})), while in the BCS approximation, let us insist, as worked out in Ref. \cite{2007Suhonen}, also contains an off diagonal terms (\ref{Eq.87}). Most likely, in order to get this off diagonal term in our formulation, we would have to resort to a perturbative treatment of the Gorkov equations.

%_____________
%_____________
\section{Conclusions and discussions}\label{sec.conclusion}
In this work we have developed, in an alternative way, the solution for the pairing Hamiltonian in the Gorkov framework. Our solution is applied to the separable pairing interaction, which allow to consider volume or surface pairing, through the coefficient of the interaction. We have got a close form for the ground state energy. It could be of major interest, to compare this result with that of the exact one, given by Pan and collaborators. The comparison with exact soluble models shows that the solution improves with the degeneracy of the system and with the inclusion of the off diagonal correlations. It is our plan to develop a perturbative solution of the Gorkov equations, to incorporate more correlations.  The next step will be to include the continuum spectrum of energy in the algebraic Gorkov formulation to deal with finite open quantum systems.

%_____________
%_____________
% acknowledgments
\ack
During the preparation of the manuscript we benefited from clarifying conversations with Luis Manuel. This work has been supported  by the National Council of Research PIP-625, CONICET, Argentina.

%_____________
%_____________
\appendix

%_____________
\section{Appendix: Time derivative of the creation operator} \label{app.1}
These are a few equations which complement the calculation of the Eqs. (\ref{eq.dgdt}) and (\ref{eq.dfdt}) in subsection \ref{sec.eqmotion}. These equations are obtained by using the relation for the time evolution of $c_{\alpha}(t)$,
\begin{equation} \label{eq.ct}
     i\hbar \, \frac{\partial }{\partial t}  c_{\alpha}(t) 
     				= \left[  c_{\alpha}(t) , \mathcal{H} \right] 
     				= e^{\frac{i}{\hbar}\mathcal{H} t} 
     	          \left[  c_{\alpha} , \mathcal{H} \right] 
     	          e^{-\frac{i}{\hbar}\mathcal{H} t} 
\end{equation}
in terms of the commutators $[c_{\alpha}, \mathcal{H}_0 ]$ and $[c_{\alpha}, V]$,
\begin{align}
       \left[  c_{\alpha}, \mathcal{H}_0  \right] &= e_{a} c_{\alpha} \label{eq.ch0} \\
     %%%
     \left[   c_{\alpha}, V  \right] &= 
     				\frac{1}{2} \sum_{\beta \gamma \delta} \, 
         \bar{V}_{\alpha \beta \gamma \delta }
          c^\dagger_\beta  \, c_\delta \, c_\gamma \, \label{eq.cv} 
\end{align}
which gives
\begin{equation}
      i\hbar \, \frac{\partial }{\partial t}  c_{\alpha}(t) 
     				= e_{a} c_{\alpha}(t)
     	 		+  \frac{1}{2} \sum_{\beta \gamma \delta}
                     \bar{V}_{\alpha \beta \gamma \delta }
                      c^\dagger_\beta(t) \, c_\delta(t) \, c_\gamma(t) \label{eq.dcdt}    
\end{equation}

%____________________________________________________
\section{Appendix: Deduction of the equations (\ref{eq.dgdt5}) and (\ref{eq.dfdt5})  of section \ref{sec.eqmotion}.} \label{app.2}
Let us start from Eq. (\ref{eq.dgdta}),
\begin{align}
   i\hbar \,\frac{\partial }{\partial t} G_{\alpha \alpha'}(t,t') &=
           \delta(t-t') \, \delta_{\alpha \alpha'}
            + e_a G_{\alpha \alpha'}(t,t') \nonumber \\
     & - \frac{g\, f_a}{i \hbar}
       \sum_{\gamma>0} \, f_c\,
    	 \langle \Psi |  T [ c^\dagger_{\bar{\alpha}}(t) c_{\bar{\gamma}}(t) 
	 		c_{\gamma}(t) c^\dagger_{\alpha'}(t')] | \Psi \rangle \label{eq.16b}
\end{align}

The average of the product of the above four operators is treated in the factorized mean-field approximation \cite{1958Gorkov,1971Fetter},
\begin{align}
   i\hbar \,\frac{\partial }{\partial t} G_{\alpha \alpha'}(t,t') &=
           \delta(t-t') \delta_{\alpha \alpha'}
            + e_a G_{\alpha \alpha'}(t,t') \nonumber \\
            & - \frac{g\, f_a}{i \hbar} \sum_{\gamma>0} \, f_c \,
         \left[
         	\langle \Psi |  T c^\dagger_{\bar{\alpha}}(t) c_{\bar{\gamma}}(t) | \Psi \rangle
         	\langle \Psi |  T c_{\gamma}(t) c^\dagger_{\alpha'}(t')] | \Psi \rangle
         	\right. \nonumber \\
      &  - \left.	
            \langle \Psi |  T c^\dagger_{\bar{\alpha}}(t) c_{\gamma}(t) | \Psi \rangle
            \langle \Psi |  T c_{\bar{\gamma}}(t)  c^\dagger_{\alpha'}(t') | \Psi \rangle
             \right. \nonumber \\ 
        & + \left.
           	   \langle \Psi | 
           	   			 T c^\dagger_{\bar{\alpha}}(t) c^\dagger_{\alpha'}(t')
           	       	| \Psi \rangle 
           	    \langle \Psi | T c_{\bar{\gamma}}(t) c_{\gamma}(t)  | \Psi \rangle
	    \right] \nonumber
%%%%%
%    \nonumber \\ 
%%%%%
\end{align}

Next we write the above equation  in term of $G$ and $\tilde{G}$ given in Eqs. (\ref{eq.g0}) and (\ref{eq.f}),
% Eqs. (\ref{eq.dgdt2}) and (\ref{eq.dfdt2}) in term of $G$ and $\tilde{G}$,
\begin{align}
   i\hbar \,\frac{\partial }{\partial t} G_{\alpha \alpha'}(t,t') &=
           \delta(t-t') \, \delta_{\alpha \alpha'}
            + e_a G_{\alpha \alpha'}(t,t')  
      - \frac{g f_a \hbar}{i} \sum_{\gamma>0} \, f_c
      \left[
         G_{\bar{\gamma} \bar{\alpha}}(t,t) G_{\gamma \alpha'}(t,t') \right.  \nonumber \\
       & \left. - G_{\gamma \bar{\alpha}}(t,t)  G_{\bar{\gamma} \alpha'}(t,t') 					
         - \tilde{G}^\dagger_{\bar{\alpha} \alpha'}(t,t') 
                \tilde{G}_{\bar{\gamma} \gamma}(t,t)
      \right] \label{eq.dgdt3}    %\\
%%%%%%
%    \nonumber \\ 
\end{align}

The first and second terms in the square bracket can be work out to give,
%in Eq. (\ref{eq.dgdt3}) can be work out to give,
\begin{align}
     & \frac{g f_a \hbar}{i} \sum_{\gamma>0} \, f_c
      \left[
         G_{\bar{\gamma} \bar{\alpha}}(t,t) G_{\gamma \alpha'}(t,t')
         - G_{\gamma \bar{\alpha}}(t,t)  G_{\bar{\gamma} \alpha'}(t,t') 
       \right] = \nonumber \\
     &   \frac{g f_a \hbar}{i} \sum_{\gamma>0} \, f_c 
      \left[ 
			\delta_{\bar{\gamma} \bar{\alpha}}
			G_{\bar{\alpha} \bar{\alpha}}(t,t)
			\delta_{\gamma \alpha'}
		    G_{\alpha' \alpha'}(t,t') 
		- \delta_{\gamma \bar{\alpha}}
		   G_{\bar{\alpha} \bar{\alpha}}(t,t)
		   \delta_{\bar{\gamma} \alpha'}
		    G_{\alpha' \alpha'}(t,t') 
       \right]_{m_\alpha>0} = \nonumber \\
     %%%%
    & 
     \frac{g f_a \hbar}{i} \sum_{\gamma>0} \, f_c   
     \delta_{\gamma \alpha}
     G_{\alpha \alpha}(t,t)
     \delta_{\gamma \alpha'}
     G_{\alpha' \alpha'}(t,t')  = \nonumber \\
     %%%%
    &   \frac{g f^2_a \hbar}{i} \,
           \delta_{\alpha \alpha'}
           G_{\alpha \alpha}(t,t)
           G_{\alpha \alpha}(t,t') 
           =
            g f^2_a  \, n_a \,
           \delta_{\alpha \alpha'} 
            G_{\alpha \alpha}(t,t') 
           =
            g f^2_a  \, n_a \,
            G_{\alpha \alpha'}(t,t') 
             \label{eq.99} %\nonumber
\end{align}
where we have used: 
(i) $G_{\alpha \beta}=\delta_{\alpha \beta} G_{\alpha \alpha}$ in the first and last equality, 
(ii) $\delta_{\bar{\gamma} \bar{\alpha}}=\delta_{\gamma \alpha}$,  
$G_{\bar{\alpha} \bar{\alpha}}=G_{\alpha \alpha}$, 
$\delta_{\gamma \bar{\alpha}}=0$, in the second equality, and 
(iii) $G_{\alpha \alpha}(t,t)=\frac{i}{\hbar} n_a$ in the last one.

Using this result we obtain the correction $- g f^2_a  \, n_a$ to the single-particle energy $\varepsilon_a$, which is beyond the Hartree-Fock contribution \cite{1966Irvine}. Then, in terms of the self-energy $\tilde{e}_a=e_a - g f^2_a  \, n_a$, the Eq. (\ref{eq.dgdt3}) reads, 
\begin{equation}
     i\hbar \,\frac{\partial }{\partial t} G_{\alpha \alpha'}(t,t') =
           \delta(t-t') \delta_{\alpha \alpha'}
            + \tilde{e}_a G_{\alpha \alpha'}(t,t')  
      + \frac{g f_a \hbar}{i} \sum_{\gamma>0} \, f_c \,
                 \tilde{G}^\dagger_{\bar{\alpha} \alpha'}(t,t') 
                \tilde{G}_{\bar{\gamma} \gamma}(t,t) \nonumber %\label{eq.dgdt4} 
\end{equation}

Finally, using the fact that the Green functions depend on the difference $\tau=t-t'$, Eq. (\ref{eq.16b}) reads,
\begin{align*}
  & \left[ i\hbar \frac{\partial }{\partial \tau}  - \tilde{e}_a  \right] G_{\alpha \alpha'}(\tau) 
      - \frac{\hbar g f_a}{i} 
         \sum_{\gamma>0} \, f_c \,
                \tilde{G}_{\bar{\gamma} \gamma}(0)
                \tilde{G}^\dagger_{\bar{\alpha} \alpha'}(\tau)  =
           \delta(\tau) \, \delta_{\alpha \alpha'}  %\\
\end{align*}
which is Eq. (\ref{eq.dgdt5}) of section \ref{sec.eqmotion}.

Similarly, starting from Eq. (\ref{eq.dfdta}),
\begin{align}
   i\hbar \,\frac{\partial }{\partial t} \tilde{G}^\dagger_{\alpha \alpha'}(t,t') &=
            - e_a \tilde{G}^\dagger_{\alpha \alpha'}(t,t') 
       + \frac{g\, f_a}{i \hbar} 
       \sum_{\gamma>0} \, f_c\,
    	 \langle \Psi |  T [ c^\dagger_{\gamma}(t) c^\dagger_{\bar{\gamma}}(t) 
	 		c_{\bar{\alpha}}(t) c^\dagger_{\alpha'}(t')] | \Psi \rangle \nonumber % \label{eq.17b}
\end{align}
the last term is factorized as
\begin{align}
 &  i\hbar \,\frac{\partial }{\partial t} \tilde{G}^\dagger_{\alpha \alpha'}(t,t') =
            - e_a \tilde{G}^\dagger_{\alpha \alpha'}(t,t') 
  + \frac{g\, f_a}{i \hbar} \sum_{\gamma>0} \, f_c \,
      \left[
         \langle \Psi | 
         		 T c^\dagger_{\gamma}(t) c^\dagger_{\bar{\gamma}}(t)  	| \Psi \rangle 
         \langle \Psi |  T c_{\bar{\alpha}}(t) c^\dagger_{\alpha'}(t') 
         | \Psi \rangle  \right.  \nonumber \\
 &   - \left.	 
           \langle \Psi |  T c^\dagger_{\gamma}(t) c_{\bar{\alpha}}(t)  | \Psi \rangle 
            \langle \Psi | T c^\dagger_{\bar{\gamma}}(t) c^\dagger_{\alpha'}(t')
               | \Psi \rangle   \right.  
       +   \left.
             \langle \Psi | T c^\dagger_{\gamma}(t)
             									c^\dagger_{\alpha'}(t')  | \Psi \rangle 
                \langle \Psi | T c^\dagger_{\bar{\gamma}}(t) c_{\bar{\alpha}}(t) 	 | \Psi \rangle			 \right] \nonumber % \label{eq.dfdt2}
\end{align}
which, in terms of $G$ and $\tilde{G}$ reads,
\begin{align}
   i\hbar \,\frac{\partial }{\partial t} \tilde{G}^\dagger_{\alpha \alpha'}(t,t') &=
            - e_a \tilde{G}^\dagger_{\alpha \alpha'}(t,t')  
  + \frac{g f_a \hbar}{i} \sum_{\gamma>0} \, f_c
       \left[
           - \tilde{G}^\dagger_{\gamma \bar{\gamma}}(t,t) 
           					G_{\bar{\alpha} \alpha'}(t,t') \right.  \nonumber \\
             & \left.  - G_{\bar{\alpha} \gamma}(t,t) 
           				\tilde{G}^\dagger_{\bar{\gamma} \alpha'}(t,t')
           + \tilde{G}^\dagger_{\gamma \alpha'}(t,t') 
           				G_{\bar{\alpha} \bar{\gamma}}(t,t)  \right]  \nonumber % \label{eq.dfdt3}
\end{align}

The sum of the second and third terms of the square bracket gives,
\begin{equation}
      \frac{g f_a \hbar}{i} \sum_{\gamma>0} \, f_c
       \left[
           - G_{\bar{\alpha} \gamma}(t,t) 
           				\tilde{G}^\dagger_{\bar{\gamma} \alpha'}(t,t')
           + \tilde{G}^\dagger_{\gamma \alpha'}(t,t') 
           				G_{\bar{\alpha} \bar{\gamma}}(t,t)  \right] 
   =g\, f^2_a \, n_a\, \tilde{G}^\dagger_{\alpha \alpha'}(t,t')   
\end{equation}
which gives the same correction, now for the term $ \tilde{G}^\dagger_{\alpha \alpha'}(t,t')$. 

Finally, in terms of $\tau$, the equation of motion reads,
\begin{align}
  &  \left[  i\hbar \frac{\partial }{\partial \tau} + \tilde{e}_a \right] 
   				\tilde{G}^\dagger_{\alpha \alpha'}(\tau)  
  + \frac{\hbar g f_a}{i}  
      \sum_{\gamma>0} \, f_c \,
           \tilde{G}^\dagger_{\gamma \bar{\gamma}}(0) 
           				G_{\bar{\alpha} \alpha'}(\tau)   =0      
\end{align}
which is Eq. (\ref{eq.dfdt5}) of section \ref{sec.eqmotion}.

%_________________________________________
\section{Appendix: Evaluation of the anomalous Gorkov equation to equal time} \label{app.3}
In this subsection we show that both limits $\tau \rightarrow 0^\pm$ of $\tilde{G}^\dagger_{\gamma \bar{\gamma}}(\tau)$ give the same result. Let us start from the Fourier transform 
\begin{equation} \label{eq.tf} 
   \tilde{G}^\dagger_{\gamma \bar{\gamma}}(\tau) =
        \int_{-\infty}^\infty  
        \tilde{G}^\dagger_{\gamma \bar{\gamma}}(E) 
        e^{-i\frac{E}{\hbar}\tau}
         \frac{dE}{2\pi \hbar} 
\end{equation}

One of the limits reads,
\begin{align}
    \tilde{G}^\dagger_{\gamma \bar{\gamma}}(0^+) &=
       \lim_{\tau \rightarrow 0^+} 
       \int_{-\infty}^\infty  
        \tilde{G}^\dagger_{\gamma \bar{\gamma}}(E) 
        e^{-i\frac{E}{\hbar}\tau}
         \frac{dE}{2\pi \hbar} 
\end{align}
in order the integral to converge we have to close the contour in the lower complex energy plane. In this way the pole is at $E=E_c - i |\eta|$, then
\begin{align}
    \tilde{G}^\dagger_{\gamma \bar{\gamma}}(0^+) &=
       \lim_{\tau \rightarrow 0^+} 
       \frac{(-2 \pi i)}{2\pi \hbar}  
       \left( - \frac{\Delta^*_c}{2E_c} \right)
        =  \frac{i}{ \hbar} \frac{\Delta^*_c}{2E_c}  \label{eq.f3}
\end{align}
where we used 
$\textrm{Res}[\tilde{G}^\dagger_{\gamma \bar{\gamma}}(E=E_c - i |\eta|)]=- \frac{\Delta^*_c}{2E_c}$.

Alternatively, we may calculate 
\begin{align}
    \tilde{G}^\dagger_{\gamma \bar{\gamma}}(0^-) &=
       \lim_{\tau \rightarrow 0^-} 
       \int_{-\infty}^\infty  
        \tilde{G}^\dagger_{\gamma \bar{\gamma}}(E) 
        e^{-i\frac{E}{\hbar}\tau}
         \frac{dE}{2\pi \hbar} 
\end{align}
in this case the contour has to be closed in the upper part of the complex $E$ plane to give
\begin{align}
    \tilde{G}^\dagger_{\gamma \bar{\gamma}}(0^-) &=
       \lim_{\tau \rightarrow 0^-} 
       \frac{(2 \pi i)}{2\pi \hbar}  
        \frac{\Delta^*_c}{2E_c}
        =  \frac{i}{ \hbar} \frac{\Delta^*_c}{2E_c} \label{eq.f4}
\end{align}

Equations (\ref{eq.f3}) and (\ref{eq.f4}) shows that
\begin{equation*}
    \tilde{G}^\dagger_{\gamma \bar{\gamma}}(0) 
        =  \frac{i}{ \hbar} \frac{\Delta^*_c}{2E_c} 
\end{equation*}

%_______________________________________
\section{Appendix: Spectral representation of the model energy} \label{app.4}
The modified energy $\mathcal{E}$ is calculated using the hole part of the spectral function calculated from the Green function. In the process, a closure relation in the Fock space is used as an intermediate step. We followed closely Ref. \cite{2005Dickhoff}.

Let us start from the following integral
\begin{equation} \label{eq.einitial}
   I_\alpha = \frac{1}{\pi} \int_{-\infty}^{\lambda} \, E \,
   		\Im(G_{\alpha \alpha}(E)) \, dE
\end{equation}
where $\Im(G_{\alpha \alpha}(E))$ is the imaginary part of $G_{\alpha \alpha}(E)$.

The Fourier transform of $G_{\alpha \alpha'}(E)$ reads
\begin{align}
   G_{\alpha \alpha'}(E) &= \int_{-\infty}^\infty \, \frac{d(t-t')}{i \hbar} \, 
   			e^{\frac{i}{\hbar} E (t-t')}
       \left[
          \theta(t-t') \sum_n 
          			\langle \Psi | c_\alpha(t) | \Psi_n \rangle
          			\langle \Psi_n | c^\dagger_{\alpha'}(t') | \Psi \rangle \right. \nonumber \\
          & \left.	-
           \theta(t'-t) \sum_m			
           			\langle \Psi | c^\dagger_{\alpha'}(t') | \Psi_m \rangle
           			\langle \Psi_m | c_\alpha(t) | \Psi \rangle
       \right] \\
  %%%%
   &= \int_{-\infty}^\infty \, \frac{d\tau}{i \hbar}
       \left[
          \theta(\tau) 
             \sum_n \, e^{\frac{i}{\hbar} \tau (E+\mathcal{E}-\mathcal{E}_n)}
          			| \langle \Psi_n | c^\dagger_\alpha | \Psi \rangle |^2 \right. \nonumber \\
         & \left.	   	-
           \theta(-\tau) 
              \sum_m \, e^{\frac{i}{\hbar} \tau (E+\mathcal{E}-\mathcal{E}_m)}			
           			| \langle \Psi_m | c_{\alpha'} | \Psi \rangle |^2
       \right]      \\
     %%%%
   &= \int_{-\infty}^\infty \, \frac{d\tau}{i \hbar} \,
       \left[
				- \sum_n \, 
				 \int_{-\infty}^\infty \, \frac{dE'}{2\pi i} \, 
				 	\frac{e^{\frac{i}{\hbar} \tau (-E'+E+\mathcal{E}-\mathcal{E}_n)}}
				 						 {E'+i\eta}
				 	| \langle \Psi_n | c^\dagger_\alpha | \Psi \rangle |^2		\right. \nonumber \\			 
			  & \left.		 +
				  \sum_m \, 
				   \int_{-\infty}^\infty \,\frac{dE'}{2\pi i} \, 
				   \frac{e^{\frac{i}{\hbar} \tau (E'+E+\mathcal{E}-\mathcal{E}_m)}	}
				   		 {E'+i\eta}
				   	| \langle \Psi_m | c_{\alpha'} | \Psi \rangle |^2
       \right] \nonumber   \\
         %%%%
   &= \int_{-\infty}^\infty  \,  \frac{ dE'}{2\pi \hbar}    
   			 \left[
				 \sum_n \, \frac{2\pi \hbar \, \delta(-E'+E+\mathcal{E}-\mathcal{E}_n)}{E'+i\eta}
						| \langle \Psi_n | c^\dagger_\alpha | \Psi \rangle |^2 \right. \nonumber \\	
			  & \left.	-
				 \sum_m \, \frac{2\pi \hbar \, \delta(E'+E+\mathcal{E}-\mathcal{E}_m)}{E'+i\eta}
						| \langle \Psi_m | c_{\alpha'} | \Psi \rangle |^2
			 \right]	 \nonumber \\
         %%%%
   &=     \sum_n \, \frac{| \langle \Psi_n | c^\dagger_\alpha | \Psi \rangle |^2}
										{E+\mathcal{E}-\mathcal{E}_n+i\eta}		
  			    -
  			     \sum_m \, \frac{ | \langle \Psi_m | c_{\alpha'} | \Psi \rangle |^2}
  			     				{-E-\mathcal{E}+\mathcal{E}_m+i\eta}		 
\end{align}
in the first equality we used 
$\int d(t-t') G_{\alpha \alpha'}(t-t') e^{\frac{i}{\hbar} E (t-t')}=G_{\alpha \alpha'}(E)$ and we introduced the closure relation in the Fock space 
$\sum_{n(m)} | \Psi_{n(m)} \rangle \langle \Psi_{n(m)} |$, with 
$\langle \Psi_{n(m)} | \mathcal{H} | \Psi_{n'(m')} \rangle = \mathcal{E}_{n(m)} \delta_{n(m),n'(m')}$. In the second equality we introduced $\tau=t-t'$. In the third equality we used the identity 
$\theta(\tau)=-\int \frac{dE'}{2\pi i} \frac{e^{-\frac{i}{\hbar}\tau E'}}{E'+i\eta}$ with $\eta$ a positive infinitesimal.

The imaginary part is obtained using
$\int d\tau e^{\frac{i}{\hbar} E \tau} = 2\pi \hbar \delta(E)$, and
$\lim_{\eta \rightarrow 0} \frac{\eta}{E^2-\eta^2}=\pi \delta(E)$,
\begin{align}
   \Im G_{\alpha \alpha'}(E)
   &=  -  \pi \sum_n \, \delta(E+\mathcal{E}-\mathcal{E}_n)
						| \langle \Psi_n | c^\dagger_\alpha | \Psi \rangle |^2 \nonumber \\
  	&	   +
  			  \pi   \sum_m \,  \delta(E-\mathcal{E}+\mathcal{E}_m)
  			           | \langle \Psi_m | c_{\alpha'} | \Psi \rangle |^2			 
\end{align}
The difference with the usual particle or hole \cite{1992Rijsdijk} spectral function, is that here, they represent the probability for adding or removing a single nucleon from a Fock ground state with and average number $A$ of particles, while ending up in another Fock excited (or ground) state with average $A\pm1$ particles.

Then, the magnitude $I_\alpha$ takes only the hole part of $G$,
\begin{equation}
   I_\alpha = \frac{1}{\pi} \int_{-\infty}^\lambda \, dE \, E \, \Im G_{\alpha \alpha}(E)
   				=  \sum_m \,   (\mathcal{E}-\mathcal{E}_m)
  			           | \langle \Psi_m | c_{\alpha} | \Psi \rangle |^2		
\end{equation}

The, next step is to make the r.h.s. independent of the many-body basis $| \Psi_m \rangle$, and this is achieved by taking out the closure relationship
\begin{align}
     I_\alpha &=   \sum_m \,   (\mathcal{E}-\mathcal{E}_m)
  			            \langle \Psi | c^\dagger_{\alpha} | \Psi_m \rangle
  			             \langle \Psi_m | c_{\alpha} | \Psi \rangle  \\
  	%%%%
  	 &=   \sum_m \, 
  			            \langle \Psi |  c^\dagger_{\alpha}  | \Psi_m \rangle
  			             \langle \Psi_m | c_{\alpha} \mathcal{H} | \Psi \rangle  		             
  	      - 
  	      \sum_m \, 
  			            \langle \Psi | c^\dagger_{\alpha} \mathcal{H} | \Psi_m \rangle
  			             \langle \Psi_m | c_{\alpha} | \Psi \rangle   \nonumber		\\
  		%%%%
  	 &=   
  			            \langle \Psi | c^\dagger_{\alpha} c_{\alpha} \mathcal{H} | \Psi \rangle  		             
  	      - 
  			            \langle \Psi | c^\dagger_{\alpha} \mathcal{H}  c_{\alpha} | \Psi \rangle   \nonumber		\\		          
  		 		%%%%
  	 &= \langle \Psi |  c^\dagger_{\alpha} [c_{\alpha},\mathcal{H}] | \Psi \rangle                           
\end{align}

One more step is needed to make appear the ground state energy. Using equations (\ref{eq.ch0}) and (\ref{eq.cv}), and summing on $\alpha$, we get 
\begin{equation}
   \sum_\alpha I_\alpha = 
         \langle \Psi | \mathcal{H}_0  | \Psi \rangle 
           +2  \langle \Psi | V  | \Psi \rangle 
\end{equation}

Then, we can write the ground state energy in terms of the Green function and the diagonal part of 
$\langle \Psi | \mathcal{H}  | \Psi \rangle= \langle \Psi | \mathcal{H}_0  | \Psi \rangle + \langle \Psi | V | \Psi \rangle $,
\begin{equation} \label{eq.egs}
   \mathcal{E} = \frac{1}{2}  \langle \Psi | \mathcal{H}_0  | \Psi \rangle 
   					+  \frac{1}{2} \sum_\alpha I_\alpha 
\end{equation}
with
\begin{align}
   \langle \Psi | \mathcal{H}_0  | \Psi \rangle &= \sum_\alpha \, v^2_a e_a  \\
    \sum_\alpha I_\alpha &=
     \frac{1}{\pi} \sum_\alpha  \int_{-\infty}^\lambda \, dE \, E \, \Im G_{\alpha \alpha}(E)
\end{align}
where we have used 
$\langle \Psi | e_a c^\dagger_\alpha c_\alpha  | \Psi \rangle = e_a n_\alpha = e_a v^2_a$. 

The imaginary part of the Green function can be calculated explicitly
\begin{align}
    \Im G_{\alpha \alpha}(E) &= \Im \frac{u_a^2}{E-E_a+i |\eta|}
    									+ \Im \frac{v_a^2}{E+E_a-i |\eta|} \\
   %%%%
    &=  - u_a^2 \frac{|\eta|}{(E-E_a)^2+ \eta^2}
    									+ v_a^2 \frac{|\eta|}{(E+E_a)+ \eta^2}  	\\								
   %%%%
    &=  - u_a^2 \pi \delta(E-E_a) + v_a^2 \pi \delta(E+E_a)								
\end{align}
in the last term the limit $\lim_{|\eta| \rightarrow 0}$ was taken.

Then,
\begin{align}
    \sum_\alpha I_\alpha &=
     \sum_\alpha  \int_{-\infty}^\lambda \, dE \, E \, 
     \left[ 
              - u_a^2 \delta(E-E_a) + v_a^2 \delta(E+E_a)
     \right] \nonumber \\
     %%%%
     &= - \sum_\alpha \, E_a \,  v_a^2  \label{eq.efinal2}
\end{align}

%__________
% Bibliography
%\bibliographystyle{elsarticle-num} 
%\bibliography{bcs2018}

\end{document}